\documentclass[pra,twocolumn,showpacs,superscriptaddress,floatfix]{revtex4}

\usepackage{graphicx}

\begin{document}
\title{Interaction-driven effects on two-component Bose-Einstein
condensates }
\author{D. M. Jezek}
\affiliation{Departamento de F\'{\i}sica, Facultad de Ciencias Exactas
y Naturales, \\ Universidad de Buenos Aires, RA-1428 Buenos Aires,
Argentina}
\affiliation{Consejo Nacional de Investigaciones Cient\'{\i}ficas y
T\'ecnicas, Argentina } \author{P. Capuzzi} \affiliation{Departamento
de F\'{\i}sica, Facultad de Ciencias Exactas y Naturales, \\
Universidad de Buenos Aires, RA-1428 Buenos Aires, Argentina}

\begin{abstract}

We investigate the role of the interparticle-interaction strength in
the distribution of two species of atoms inside a condensate.  We
focus upon the study of systems for which the minima of the trapping
potentials for the species are displaced from each other by a distance
that is small compared to the size of the total condensate. We show
that in a small range of the interparticle-interaction strength the
distribution of species undergoes dramatic changes, and exhibits a
variety of different features.  We demonstrate that this behavior can
be easily understood in terms of the Thomas-Fermi approximation.  This
effect may be useful in experimentally determining the values of the
scattering lengths.

\end{abstract}
\date{\today}
\pacs{03.75. Fi, 05.30.Jp, 67.90.+z,32.80.Pj }

\maketitle Much experimental and theoretical work in Bose-Einstein
Condensation deals with systems composed of a mixture of two distinct
species of atoms \cite{fe01,ma98,ma99,an00,an01,ha98}.  For example,
one mixture commonly used is that of atoms of $^{87}$Rb in two
different hyperfine states $ |F=1, m_f=-1\rangle $ and $ |F=2,
m_f=1\rangle $ \cite{ma98}.  This mixture has shown itself to be very
useful for experimentally generate topological modes, such as
different types of vortices \cite{ma99,an00,an01}.  Experimentally
these mixtures are under conditions such that the species behave as
``effectively distinguishable", and have been observed to separate
partially in space \cite{ha98}.  An interesting issue to investigate
is how the particles distribute themselves inside the condensate
depending on the interparticle interaction, specially because there
still exists uncertainty about its numerical value.

From a theoretical point of view, H. Pu and N. P. Bigelow studied the
 ground state properties of different mixtures in a spherically
 symmetric trapping potential.  In Ref.  \cite{pu98}, they displayed
 the density profiles of both components for various combinations of
 scattering lengths below phase separation, testing the validity of
 the Thomas- Fermi approximation.  Later, E. Timmermans extended the
 study to phase separation \cite{ti98}.  In particular he showed that
 in a spherically symmetric trap the less repulsive component remains
 inside a sphere while the other component lies in a spherical shell
 around it.

In this work we analyze the case when one of the species is in a
slightly shifted potential with respect to the other species, and thus
the spherical symmetry is broken.  We find that by varying the
interparticle interaction strength by less than $ 10$\%, the particles
rearrange themselves in very different configurations inside the
condensate.  As far as we know, this problem was only studied for a
particular set of scattering lengths \cite{es97} and it still lacks an
easy interpretation.  Here we show that the distribution of species
can be easily understood in terms of the Thomas-Fermi (TF)
approximation.  We compute the exact Gross-Pitaevskii (GP) solutions
for a large number of particles and find that TF predictions describe
with great accuracy the geometrical properties of the distribution of
species within the condensate.
 
In order to describe the wave functions of two-species condensates one
has to solve the coupled Gross-Pitaevskii equations \cite{da99}:
\begin{eqnarray}
 \left (-\frac{ \hbar^2 \nabla^2}{2 M_i}+ V_i 
+ N_i G_{i,i} |\Psi_i|^2 \right. 
 & + &  \left.  \sum_k N_k G_{i,k}  |\Psi_k|^2 \right ) \, \Psi_i
  \nonumber \\ 
  &  = &    \mu_i \Psi_i.
\label{gp}
\end{eqnarray}
where $ N_i$ and $ M_i$ denote the number of atoms and the mass of the
species $ i $, respectively.  The factors $G_{k,l}= u_{k,l} U $, with
$ u_{k,l} $ being the relative interaction strengths between species
$k$ and $l$, and $ V_i $ the potential seen by species $i $.
Hereafter, we consider a two component system and set the most
repulsive component $|1\rangle$ in the $ \Psi_1 $ state, fixing $
u_{1,1}= 1$ $ (> u_{2,2}) $, so $ U= 4 \pi \hbar^2 a / M_1 $, $ a $
being the scattering length of the $|1\rangle$-species.  In addition,
for simplicity, we set $M_1=M_2=M $.

\paragraph*{Qualitative analysis---}

In order to study qualitatively how the species rearrange themselves
inside the condensate when varying the relative interparticle strength
$ u_{1,2}$, we make use of the Thomas Fermi (TF) approximation, which
consists of neglecting the kinetic energy, and thus removing the
laplacian terms in Eqs. (\ref{gp}).  We will consider a system in
which both components are in an axially symmetric trap and the $
|2\rangle $ component has the minimum shifted in the $z$ direction by
a value $ -d $.  We make a change of variables according to
$\sqrt{(M/2)} \omega_z \vec{r} \rightarrow \vec{r} $, where $
\omega_r$ and $ \omega_z $ are the trap angular frequencies in the $r$
and $z$ coordinate, respectively.  Futhermore, by defining the aspect
ratio $ \lambda = \omega_r / \omega_z $, the potentials $V_i$, written
in cylindrical variables, read
\begin{equation}
V_{1} =  \lambda^2  r^2 +  z^2
\label{pot1}
\end{equation}
and
\begin{equation}
V_{2} = \lambda^2  r^2 +   (z+d)^2.
\label{pot2}
\end{equation}

The sign of the determinant $ \Delta \equiv u_{1,1} u_{2,2} -
 u_{1,2}^2 $ defines two different features for the distribution of
 particles.  When $ \Delta > 0 $ and $ u_{1,2}$ is small, there exists
 a large coexistence region. As $ u_{1,2}$ is increased this region
 decreases. At $ \Delta = 0 $, a phase separation takes place reducing
 the coexistence region to an interface.  We shall consider the cases
 $ \Delta > 0 $ and $ \Delta < 0 $ separately.

\paragraph*{ Coexistence 
 $ \Delta > 0 $---}

 The solution of Eqs. (\ref{gp}) in the TF approximation can be easily
obtained and has the following different expressions depending on
whether there exists any overlap between the wave functions of both
species.

a) In the region where only one wave function is non vanishing ( $
|\Psi_{i}|^2 \ne 0 $ and $ |\Psi_{k}|^2 = 0 $, for $ i \ne k$) the TF
equations are decoupled and the solution reads
\begin{eqnarray}
 |\Psi_{1}|^2 & = &  \left[ \mu_1 - \lambda^2 r^2-
z^2 \right]/(G_{1,1} N_1), \nonumber \\
 |\Psi_{2}|^2 & = & \left[ \mu_2 - \lambda^2 r^2
-(z+d)^2 \right]/(G_{2,2} N_2).
\label{fua}
\end{eqnarray}

b) In the region where both wave functions are non vanishing $
 |\Psi_{i}|^2 > 0 $, the solution may be written after some algebra
 as:

\begin{eqnarray}
|\Psi_{1}|^2 & = & \left[ R_1^2  - \lambda^2  r^2
-  \left ( z- \frac{\beta_1 d }{1 - \beta_1}\right)^2 
 \right]  B_1 \,  u_{2,2}, \nonumber \\
 |\Psi_{2}|^2 & = & \left[ R_2^2  - \lambda^2  r^2
-  \left  (z+ \frac{d}{1 - \beta_2} \right)^2 
 \right] B_2 \, u_{1,1},   
\label{fu}
\end{eqnarray}
with
\begin{eqnarray}
 R_1^2 &=&  \frac{\mu_1- \beta_1 \mu_2}{(1 - \beta_1)} + 
 \frac{\beta_1 d^2 }{(1 - \beta_1)^2}, \nonumber \\
R_2^2 &=&  \frac{\mu_2- \beta_2 \mu_1}{ (1- \beta_2 )}
+ \frac{ \beta_2 d^2}{(1 - \beta_2)^2},
\label{ra}
\end{eqnarray}
where $ \beta_1 = u_{1,2} / u_{2,2} $, $ \beta_2 = u_{1,2} / u_{1,1}$
 and $ B_i = (1-\beta_i)/(U N_i \Delta ) $.

The surfaces $ \cal{S}$$_1$ and $ \cal{S}$$_2$, defined by equating
the expression inside the square bracket in $|\Psi_{1}|^2$ and $
|\Psi_{2}|^2$ of Eq. (\ref{fu}) respectively to zero, determine the
boundary of the coexistence region.  It may be seen that these
surfaces are ellipsoids centered in $ d_1= \beta_1 d / (1 - \beta_1)$
and $d_2= - d/ (1 - \beta_2) $ along the $z$-axis, respectively.  It
is worth mentioning that this result does not depend on any other
quantity, as for example the number of particles or the frequencies of
the trapping potential.  The factor $ 1 - \beta_i $ in the denominator
makes these displacements diverge when $ \beta_i $ is close to unity,
and one can guess that for these values, some dramatic effects in the
redistribution of particles could take place, as we shall discuss
later.

\paragraph*{ Phase separation  $ \Delta < 0 $ ---}

For these interaction strengths, in the Thomas-Fermi approximation,
the species are confined to two separated regions within the
condensate.  The boundary surface $ \cal{S}$$_s$ between these two
regions may be obtained by equating the pressure \cite{da99}

\begin{equation}
 P_{i}= \frac{ G_{i,i}}{2}  |\Psi_{i}|^4 
\label{p}
\end{equation}
on both sides of the interface, yielding

\begin{equation}
 |\Psi_{1}|^2  = \sqrt{\frac{ G_{2,2}}{ G_{1,1}}}  |\Psi_{2}|^2.
\label{ec}
\end{equation}
Assuming that the wave functions on each side  are given by the expression
(\ref{fua})
and defining $ a =  \sqrt{G_{2,2} /  G_{1,1}} $, the 
interface  obtained is 

\begin{equation}
 R_s^2 - \lambda^2 r^2 - \left (z+ \frac{d}{1 - a}\right)^2 = 0
\label{sur}
\end{equation}
with

\begin{equation}
 R_s^2 = \frac{\mu_2- a \mu_1}{ (1- a )}
+ \frac{a d^2}{(1 - a)^2}. 
\label{rs}
\end{equation}

This surface is an ellipsoid whose center is displaced in $ d_s = -d/
(1-a) $ along the $z$-axis. Note that the quantities $R_s$ and $d_s$
do not depends on $ u_{1,2}$.  It is easy to prove that for $ \Delta
=0 $ the surfaces verify $ \cal{S}$$_1 \equiv \cal{S}$$_2 \equiv
\cal{S}$$_s$.

\begin{figure}
\centering
\includegraphics[width=8cm,clip=true]{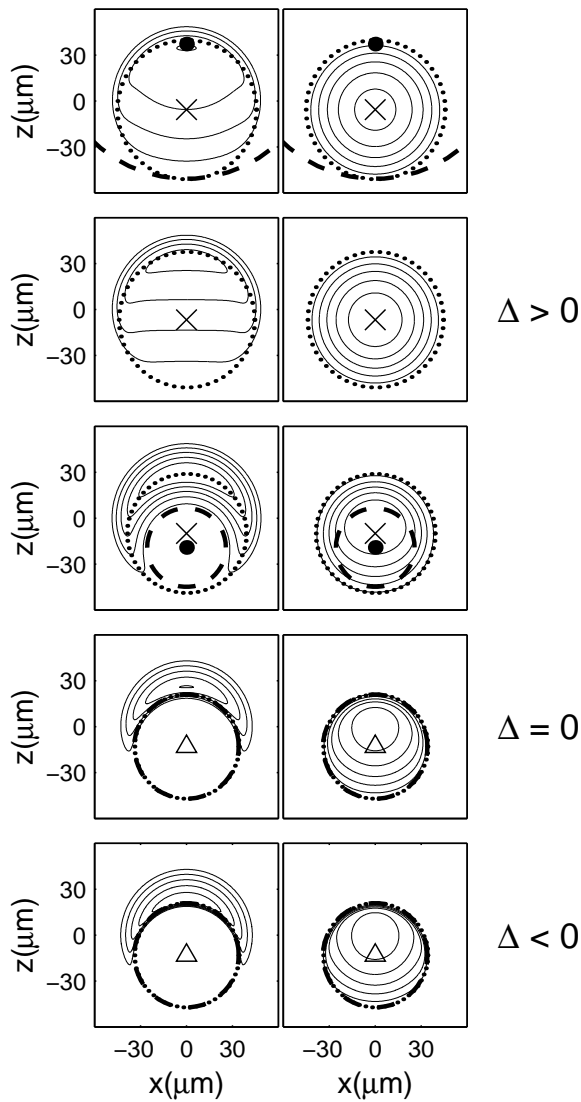}
\caption{\label{fig1} GP density contours of $\rho_1$ and $\rho_2$ are
shown in the first and second column, respectively. From top to bottom
of the graph, the rows correspond to relative interparticle
interaction strengths $ u_{1,2}$ = $0.93$, $0.94$, $0.96$, $
\sqrt{0.94}$ and $1$. The dashed, dotted and dash-dotted lines
correspond to TF radii $ R_1$, $R_2$, and $R_s$, respectively. The
dot, cross and triangle indicate $d_1$, $d_2$, and $d_s$,
respectively. The trapping potential displacement is $ d= 0.4 \mu$m.}
\end{figure} 

In what follows we consider only $ \lambda =1 $ in which case the
surfaces $ \cal{S}$$_i$ are spheres with radii $R_i$.

In the TF approximation, the density contours $ \rho_i = |\Psi_{i}|^2
  $ $(i=1,2)$ inside the coexistence region are spherical surfaces
  with radii $ R < R_i $ ( $ R > R_i $) if $\beta_i <1$ ($\beta_i >1$)
  centered in $d_i$.

\paragraph*{ Numerical results---}

 On the one hand, we compute the displacements $d_i$ and radii $R_i$
in the Thomas-Fermi approximation.  Note that these radii depend on
the chemical potentials, and thus also on both the number of particles
and the trap frequencies.  On the other hand, in order to obtain the
exact densities, we solved the Gross-Pitaevskii equations, using a
steepest descent method.

In particular, we used the relative intraparticle interaction strength
of atoms of $^{87}$Rb in the two different hyperfine states given,
within a $ 1.2 $\% of error, in Ref. \cite{ma98}, $ u_{1,1}=1 $ and $
u_{2,2}= 0.94 $.  We chose a trapping potential with an angular
frequency $ \omega_{r} = 2 \pi \times 7.8 $ Hz, and set the number of
particles of each species $N_1= N_2 = 1$$ \times$$ 10^7 $.

In Fig. \ref{fig1} we show the Gross-Pitaevskii density contours $
 \rho_i $ in the $ y=0 $ plane, together with the Thomas-Fermi
 displacements $d_i$ and radii $R_i$, for $ d= 0.4$ $ \mu$m. From the
 first to the last row we vary $ u_{1,2} $ by less than $ 10$\%.  For
 $ u_{1,2}= 0.93 $, which corresponds to the first row of the graph,
 it can be seen that a large coexistence region still exists. This
 region is given by the intersection of the circles defined by $ R_i$
 and $ d_i$, which leaves outside only a little region on the top of
 the condensate filled with $ |1\rangle $ type particles.  It can also
 be seen that inside the coexistence region, the GP density contours
 are well-described by spherical surfaces centered in the points $
 d_1$ and $d_2$ for $ \rho_1 $ and $ \rho_2 $ respectively, as
 predicted with the TF approximation.  For the second row we consider
 $ u_{1,2}= 0.94 $, and with this value $d_1$ and $R_1$ diverge.  The
 density contours inside the coexistence region are quite planar
 surfaces for the $ |1\rangle $ component, as expected from the TF
 analysis because of the above-mentioned divergences.  For $ u_{1,2}=
 0.96 $, which is displayed in the third row, three phases already
 exist: pure $ |1\rangle $ and $ |2\rangle $ components and a
 coexistence region. Once more, the contours seem to be in agreement
 with formulae (\ref{fua}) and (\ref{fu}) with only a small departure
 in the region next to the boundaries.

  Phase separation occurs for $ u_{1,2}= \sqrt{0.94} $, which
 corresponds to the fourth row, and thus the coexistence region is
 reduced to a surface. It is evident that although the GP solutions
 exhibit some overlap the general feature of the condensate is
 well-described by the TF approximation. Of course, if we take a
 smaller number of particles into account, the overlap will be
 greater.  The last row corresponds to $ u_{1,2}= 1 $ which is quite
 similar to the previous figure. This is due to the fact that formulae
 (\ref{sur}) and (\ref{rs}) do not depend on $ u_{1,2} $, as we stated
 before.

\begin{figure}
\centering
\includegraphics[width=8cm,clip=true]{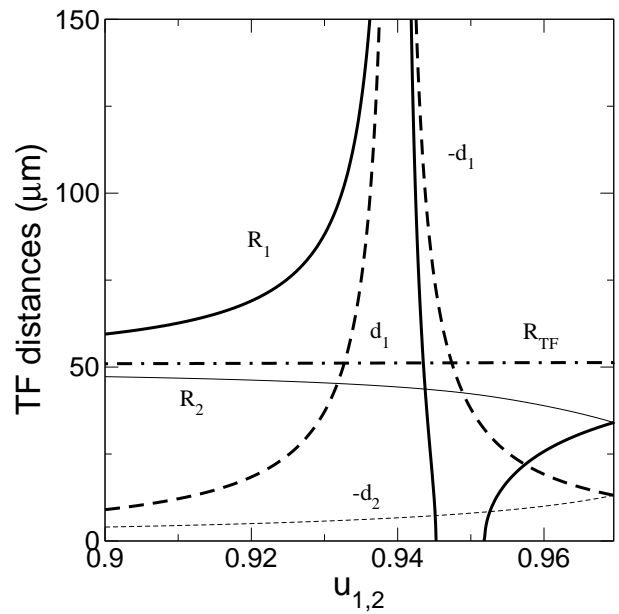}
\caption{\label{fig2} Thomas-Fermi displacements $ |d_1| $ (thick
dashed line), $ |d_2| $ (thin dashed line), radii $R_1$ (thick solid
line) and $R_2$ (thin solid line), and the TF radius of the condensate
$R_{\text{TF}}$ (dash-dotted line) as function of the interparticle
interaction strength $ u_{1,2} $ for the same conditions as in
Fig. 1.}
\end{figure}

In Fig. \ref{fig2} we display the TF quantities, $d_i$ and $ R_i$
together with the radius of the condensate $R_{\text{TF}}$ as
functions of $ u_{1,2} $. For $ u_{1,2} < 0.9 $, all these quantities
are smooth and monotonous functions, while in the interval displayed
in the figure, the quantities related to $ |1\rangle $ species exhibit
an abrupt behavior.  This effect suggests that if one wants to test
experimentally the validity of the values of the interaction strengths
within this interval by determining the displacements $ d_i $, the
error in their determination should not affect the desired quantities
too much.

\begin{figure}
\centering
\includegraphics[width=8cm,clip=true]{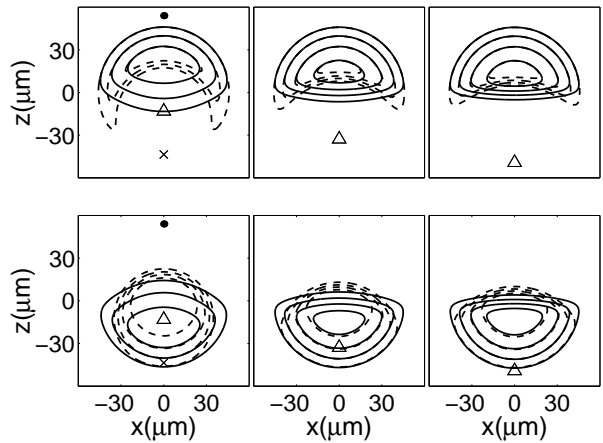}
\caption{\label{fig3} The first and second row show the GP density
contours of $\rho_1$ and $\rho_2$, respectively.  From the first to
the third column we consider $ d= 0.4$, $1$, and $1.5$ $\mu$m.  The
dashed lines correspond to the set $u_{1,1}=1,u_{1,2}= 0.97 $ and
$u_{2,2}= 0.94 $ while the solid lines correspond to the relative
interaction strengths $u_{1,1}=1, u_{1,2}= 108/109 $ and $u_{2,2}=
108.8/109 $.  The TF points $d_s$ (for the first set of parameters),
and $d_1$ and $d_2$ are marked with a triangle, a dot and a cross,
respectively.  }
\end{figure}

As a final illustration, in Fig. \ref{fig3} we display the density
contour for two different sets of relative scattering lengths to
describe the interaction of the same two species of $^{87}$Rb.  One
set, which is assumed to be more accurate \cite{ha98}, is
$u_{1,1}=1,u_{1,2}= 0.97 $ and $u_{2,2}= 0.94 $. However, to tell the
truth, up to our knowledge, the error of $u_{1,2}$ is not given
anywhere. For these values of the interaction strengths the
determinant is negative ($ \Delta= -9$$ \times$$10^{-4} $) and thus
the system is phase separated, and hence the coexistence region in the
Thomas Fermi approximation is reduced to an interface. The other,
previously used, set of parameters \cite{es97} is $u_{1,1}=1, u_{1,2}=
108/109 $ and $u_{2,2}= 108.8/109 $, and verifies $ \Delta= 1.6$$
\times$$10^{-2} $. Since the determinant is positive a coexistence
region exists. Moreover, $ \beta_1 = 0.99 $ is less than $1$, and the
contours of the $ |1\rangle $ component at the coexistence region are
still concave upwards.

We used three different displacements $ d = 0.4, 1 , 1.5 $ $ \mu$m for
 the trapping potentials.  For the first set of interaction strengths
 the corresponding TF displacements are $ d_s = -13, -32, -49$ $
 \mu$m, respectively.  While for the second set, the displacements are
 $ d_1 = 54, 135, 202$ $ \mu$m and $ d_2 = -44, -109, -163 $ $\mu$m.
 It may be seen that the distribution of particles is very different
 in both cases, especially for small displacements, and this fact can
 be used for testing experimentally the values of the scattering
 lengths. It is worth mentioning that in the experimental conditions
 of Ref. \cite{ha98}, as they obtain integrated densities, one cannot
 distinguish between a phase separated system ($\Delta < 0$) and a
 system in which the components are overlapped ($\Delta > 0$).

Finally, having in mind the relations $ \beta_1= d_1/(d+d_1) $ and $
\beta_2=(d+ d_2)/d_2 $, we wondered wether having the experimental
density contours in a given plane, say $ y=0$, one can accurately
determine the scattering lengths (see definitions of $ \beta_i $) by
estimating the displacements $d_i$.  In order to answer this question
we have done the following test.  We have used the information of the
GP density contours assuming that they represent the exact
experimental data.  For a given density contour $ z (x) $ in the $y=0$
plane we computed the slope $ b = \partial z / \partial x $ at each
point $ (x,z) $ of the curve. Then we calculated the intersection
point between the line perpendicular to the  density contour and $x=0$,
which gives the center of the spherical surface. This point is $
z_c=z+ x/b $.  Plotting $z_c$ for points all over the $ |1\rangle $ ($
|2\rangle $) component contour we found well-defined plateaus at
$z_c=0$ ( $z_c=-d$) for points outside the coexistence region, and in
$ z_c= d^*_1$ ( $ z_c= d^*_2$), where $ d^*_i$ are the estimates of
the TF $d_i$.  We used this information to determine the scattering
lengths and found an error in their determination of about $ 0.1$\%.
Note that, even considering an error in $z_c$ of about 30 $\mu$m and
using the expressions of $\beta_i$, the uncertainty in the value of
$u_{1,2}$ turns out to be below 0.5\%.  A similar procedure could be
carried out with experimental data.

In summary, we show that the distribution of components is strongly
ruled by the interparticle interaction strength, and for the number of
particles we have considered, the way the particles rearrange inside
the condensate can be easily understood in terms of the Thomas-Fermi
approximation.  We also outline a possible procedure to experimentally
determine the relative scattering lengths.

\end{document}